\documentclass[lettersize,journal]{IEEEtran}
\usepackage{amsmath,amsfonts}
\usepackage{algorithmic}
\usepackage{array}
\usepackage[caption=false,font=normalsize,labelfont=sf,textfont=sf]{subfig}
\usepackage{textcomp}
\usepackage{stfloats}
\usepackage{url}
\usepackage{verbatim}
\usepackage{graphicx}
\usepackage[linesnumbered,ruled,vlined]{algorithm2e}
\usepackage{etoolbox}\AtBeginEnvironment{algorithmic}{\small} 

\def\BibTeX{{\rm B\kern-.05em{\sc i\kern-.025em b}\kern-.08em
    T\kern-.1667em\lower.7ex\hbox{E}\kern-.125emX}}
\usepackage{balance}
\usepackage{mathtools}

\usepackage{longtable}
\usepackage{tabularx}
\usepackage{afterpage}
\usepackage{placeins}
\usepackage{rotating}
\usepackage{array}
\usepackage{todonotes}
\usepackage{xspace}
\usepackage{multirow}
\usepackage{makecell}

\newcommand{\sysold}[0]{Morphence-1.0}
\newcommand{\sysname}[0]{Morphence-2.0}
\DeclareMathOperator*{\argmax}{arg\,max}
\DeclarePairedDelimiter\ceil{\lceil}{\rceil}

\begin{document}

\title{Morphence-2.0: Evasion-Resilient Moving Target Defense Powered by Out-of-Distribution Detection \vspace*{-0.2em}}

\author{
\IEEEauthorblockN{Abderrahmen Amich, Ata Kaboudi, Birhanu Eshete}\\
 \IEEEauthorblockA{
 \textit{University of Michigan-Dearborn}\\
 \{aamich, kaboudi, birhanu\}@umich.edu}
\vspace*{-3em}}

\maketitle
\begin{abstract}
Evasion attacks against machine learning models often succeed via iterative probing of a {\em fixed target} model, whereby an attack that succeeds once will succeed repeatedly. One promising approach to counter this threat is making a model a {\em moving target} against adversarial inputs.\\ To this end, we introduce \sysname{}, a scalable moving target defense (MTD) powered by out-of-distribution (OOD) detection to defend against adversarial examples. By regularly moving the decision function of a model, \sysname{} makes it significantly challenging for repeated or correlated attacks to succeed. \sysname{} deploys a pool of models generated from a base model in a manner that introduces sufficient randomness when it responds to prediction queries. Via OOD detection,
\sysname{} is equipped with a scheduling approach that assigns adversarial examples to robust decision functions and benign samples to an undefended accurate models. To ensure repeated or correlated attacks fail, the deployed pool of models automatically expires after a query budget is reached and the model pool is seamlessly replaced by a new model pool generated in advance.\\
We evaluate \sysname{} on two benchmark image classification datasets (MNIST and CIFAR10) against 4 reference attacks (3 white-box and 1 black-box). \sysname{} consistently outperforms prior defenses while preserving accuracy on clean data and reducing attack transferability. We also show that, when powered by OOD detection, \sysname{} is able to precisely make an input-based movement of the model's decision function that leads to higher prediction accuracy on both adversarial and benign queries.
\end{abstract}


\section{Introduction}\label{sec: intro}
\IEEEPARstart{M}{achine} learning (ML) continues to propel a broad range of applications in image classification~\cite{ImageNet}, voice recognition~\cite{DL-Speech2012}, precision medicine~\cite{DeepCC2019}, malware/intrusion detection~\cite{malconv18}, autonomous vehicles~\cite{DL-autnonmous17}, and so much more. ML models are, however, vulnerable to {\em adversarial examples} ---minimally perturbed legitimate inputs that fool models to make incorrect predictions~\cite{FGSM,Biggio-ECML13}. Given an input $x$ (e.g., an image) correctly classified by a model $f$, an adversary performs a small perturbation $\delta$ and obtains $x' = x+\delta$ that is indistinguishable from $x$ to a human analyst, yet the model misclassifies $x'$. Adversarial examples pose realistic threats on domains such as self-driving cars, healthcare, and malware detection for the consequences of incorrect predictions are highly likely to cause real harm~\cite{AV-Physical-Attack17,MalConvEvade18,MalGAN17}.

To defend against adversarial examples, previous work took multiple directions each with its pros and cons. Early attempts~\cite{Early-Defense14,Early-Defense15} to harden ML models provided only marginal robustness improvements. Heuristic defenses based on {\em defensive distillation}~\cite{distillation}, {\em data transformation}~\cite{Compression17,Compression18,Augmentation17,Cropping17,Rand15,Rand18}, and {\em gradient masking}~\cite{Thermo-Encode18,PixelDefend18} were subsequently broken \cite{Carlini-Breaking17,Carlini-BreakingUsenix17,Gradient-Masking18,CW}.

While {\em adversarial training}~\cite{FGSM,EnsembelAdvTrain18} defends against known attacks, robustness comes at the expense of accuracy loss on clean data. Similarly, data transformation-based defenses also degrade accuracy on benign inputs. \textit{Certified defenses}~\cite{lecuyer2019certified,RandomSmoothing19,Certified-AdditiveNoise19} provide formal robustness guarantee, but are limited to a class of attacks constrained to LP-norms \cite{lecuyer2019certified,wong2018provable}.

As pointed out by~\cite{goodfellow2019research}, a shared limitation of prior defenses is the {\em static and fixed target} nature of the deployed ML model. We argue that, although defended by methods such as adversarial training, the fact that a ML model is a {\em fixed target} that continuously responds to prediction queries makes it {\em a prime target for repeated/correlated adversarial attacks}. As a result, given enough time, an adversary can repeatedly query the prediction API and build enough knowledge about the ML model and eventually fool it. Once the adversary launches a successful attack, it will be always effective since the model is not moving from its compromised ``location''. 

In this article, we introduce \sysname{} building on the MTD of \sysold{}~\cite{Amich_2021} in three ways: enhanced approach, more comprehensive experimental evaluations, and new insights. To differentiate advances in \sysname{}, next we first briefly summarize \sysold{} and then describe values added by \sysname{}.

\textbf{\sysold{}~\cite{Amich_2021}}. By regularly moving the decision function of a model, \sysold{}  makes it challenging for an adversary to fool the model through adversarial examples. \sysold{}  thwarts once successful and repeated attacks and attacks that succeed after iterative probing of a fixed target model through correlated sequence of attack queries. To do so, \sysold{} deploys a pool of $n$ models generated from a base model in a manner that introduces sufficient randomness when it selects the most suitable model to respond to prediction queries. The selection of the {\em most suitable model} is governed by a scheduling strategy that relies on the prediction confidence of each model on a given query input. To ensure repeated or correlated attacks fail, the deployed pool of $n$ models automatically expires after a query budget is reached. The model pool is then seamlessly replaced by a new pool of $n$ models generated and queued in advance. 
To be practical, \sysold{} aims to improve robustness to adversarial examples across white-box and black-box attacks ({\em Challenge-1}); maintain accuracy on benign samples as close to that of the base model as possible ({\em Challenge-2}); and increase diversity among models in the pool to reduce adversarial example transferability among them ({\em Challenge-3}).
It addresses {\em Challenge-1} by enhancing the MTD aspect through larger model pool size, a model selection scheduler, and dynamic pool renewal (Sections \ref{subsec:model-gen} and \ref{subsec:renewal}). {\em Challenge-2} is addressed by re-training each generated model to regain accuracy loss caused by perturbations (Section \ref{subsec:model-gen}: step-2). It addresses {\em Challenge-3} by making the individual models distant enough via distinct transformed training data used to re-train each model (Section \ref{subsec:model-gen}: step-2). Training a subset of the generated models on distinct adversarial data is an additional robustness boost to address {\em Challenges 1 and 3} (Section \ref{subsec:model-gen}: step-3).
While \sysold{} made significant advances on MTD-based prior work ~\cite{fMTD19,MTDeep19,EI-MTD} (see Section \ref{sec: related}), in this work we significantly enhance its scheduling strategy, extend experimental evaluations, and draw novel insights in the context of MTD against adversarial examples.

\textbf{\sysname{}}. We significantly overhaul the scheduling strategy in \sysold{} by introducing a layer of OOD detection that further guides the model selection strategy with respect to the nature of the received query (i.e., adversarial vs. benign). To this end, we draw insights from~\cite{OOD2IID} that makes empirical observations that suggest most adversarial examples are OOD samples. In particular, adversarial perturbations usually shift the distribution of the perturbed sample $x' = x + \delta$ away from its initial distribution $\mathbb{P}_{train}$ used in training the target model. Consequently, we leverage and adapt OOD detection~\cite{SSD} for the sake of adversarial examples detection (details in Section \ref{subsec: scheduling}).  

We evaluate \sysname{} on two benchmark image classification datasets (MNIST and CIFAR10) against four attacks: three white-box attacks (FGSM~\cite{FGSM}, PGD~\cite{PGSM}, and C\&W~\cite{CW}) and one iterative black-box attack (SPSA~\cite{uesato2018adversarial}). We compare \sysname{}'s robustness with \sysold{} and adversarial training defense of a fixed model. We then conduct detailed evaluations on the impact of the MTD strategy in defending previously successful repeated attacks and the effectiveness of OOD detection to boost the scheduling strategy. Additionally, through extensive experiments, we shed light on each component of \sysname{} and its impact towards improving the robustness results and reduce the transferability rate across models. Overall, our evaluations suggest that \sysname{} advances the state-of-the-art in robustness against adversarial examples, even in the face of strong white-box attacks such as C\&W~\cite{CW}, while maintaining accuracy on clean data and reducing attack transferability. In summary, this work builds on \sysold{} and makes the following contributions:
\begin{itemize}
\item Powered by OOD detection, \sysname{} is able to precisely select the most accurate decision function for each query.

\item \sysname{} thwarts repeated attacks that leverage previously successful attacks and correlated attacks performed through dependent consecutive queries.

\item \sysname{} outperforms adversarial training and \sysold{} while preserving accuracy on clean data and reducing attack transferability within a model pool.

\item \sysname{} improves the state-of-the-art defenses on both white-box and black-box attacks.

\item \sysname{} code is available as free and open-source software at: {\color{blue} \url{https://github.com/um-dsp/Morphence}}.
\end{itemize}

\section{Background}\label{sec: bground}

\subsection{Adversarial Examples}

 
Given a ML model $f:X \rightarrow Y$ that is trained to map an input sample $x \in X$ to a true class label $y_{true} \in Y$, $x'$ = $x  + \delta$ is called an {\em adversarial example} with an {\em adversarial perturbation} $\delta$ if: 
    $f(x') = y' \ne y_{true}, ||\delta|| < \epsilon$,
where $||.||$ is a distance metric (e.g., one of the $L_{p}$ norms)  and $\epsilon$ is the maximum allowable perturbation that results in misclassification while preserving semantic integrity of $x$. Semantic integrity is domain and/or task specific. For instance, in image classification, visual imperceptibility of $x'$ from $x$ is desired while in malware detection $x$ and $x'$ need to satisfy certain functional equivalence (e.g., both $x$  and $x'$ exhibit the same malicious behavior).
In {\em untargeted} evasion, the goal is to make the model misclassify a sample to any different class. When {\em targeted}, the goal is to make the model to misclassify a sample to a specific target class.

Adversarial examples can be crafted in {\em white-box} or {\em black-box} setting. Most gradient-based attacks~\cite{FGSM,BIM,PGSM,CW} are white-box because the adversary typically has access to model details, which allow to query the model directly to decide how to increase the model’s loss function. Gradient-based attacks assume that the adversary has access to the gradient function of the model. The goal is to find the perturbation vector $\delta^\star \in \mathbb{R}^d$ that maximizes the loss function $J(\theta, x, y_{target})$ of the model $f$, where $\theta$ are the parameters (i.e., weights) of the model $f$. In recent years, several white-box attacks have been proposed, especially for image classification tasks. Some of the most notable ones are: Fast Gradient Sign Method (FGSM)~\cite{FGSM}, Basic Iterative Method (BIM)~\cite{BIM}, Projected Gradient Descent (PGD) method~\cite{PGSM}, and Carlini \& Wagner (C\&W) method~\cite{CW}. Black-box attack techniques (e.g., MIM~\cite{MIM}, HSJA~\cite{HSJA20}, SPSA~\cite{uesato2018adversarial}) begin with initial perturbation $\delta_{0}$, and probe $f$ on a series of perturbations $x + \delta_{i}$ to craft $x'$ such that $f(x') = y' \ne y_{true}$. 
 
In this work, we use three white-box attacks (FGSM, PGD, and C\&W) and one black-box attack (SPSA). We refer the reader to \cite{Amich_2021} for technical details of these attacks.
\vspace*{-1em}
\subsection{SSD: Out-of-Distribution Detector}
\label{SSD}
SSD~\cite{SSD} is proposed to detect OOD data that lies far away from the training distribution $\mathbb{P}_{train}$ of a ML model $f$. Its appealing side is that it is a self-supervised method that can reach good performance using only unlabeled data instead of fine-grained labeled data that can be hard to produce. Given unlabeled training data, SSD leverages \textit{Contrastive self-supervised representation learning}, which aims to train a feature extractor of the in-distribution data, by discriminating between individual samples, to learn a good set of representation without the need to data labels \cite{SSD}. Next, the OOD detection is performed through a cluster-conditioned detection. Using k-means clustering,  extracted features of in-distribution data are partitioned into $m$ clusters. Features of each cluster are modeled independently to calculate an outlier score of an input $x$, using the Mahalanobis distance $M(x,\mathbb{P}_{train})$. 
It is equivalent to the euclidean distance, but scaled with eigenvalues in the eigenspace. SSD discriminates
between in-distribution (e.g CIFAR-10) and OOD (e.g CIFAR-100) data along each principal eigenvector. With euclidean distance (i.e. in absence of scaling), components with higher eigenvalues have more weights but provide least discrimination. Scaling with eigenvalues
removes the bias. In other words, $M(x,\mathbb{P}_{train}$) is more effective for outlier detection in the feature space. More details about the choice of the distance metric and the background of the Contrastive self-supervised learning can be retrieved in the cited paper \cite{SSD}. More importantly, we choose SSD for our approach, given that it outperforms all other OOD detection tools, at the time of the submission of this article.

\vspace{-1em}
 \section{Related Work}\label{sec: related}
We review related work along three lines: best-effort heuristic defenses, certified defenses, and moving target defenses.

\textbf{Best-Effort Heuristic Defenses}.
Several heuristic-based defense approaches have been proposed, most of which were subsequently broken \cite{Carlini-Breaking17,Carlini-BreakingUsenix17,Gradient-Masking18}. Many of the early attempts~\cite{Early-Defense14,Early-Defense15} to harden ML models provided only marginal robustness improvements. In the following, we highlight this line of defenses.

{\em Defensive distillation} by Papernot et al.~\cite{distillation} is a strategy to distill knowledge from neural network as soft labels by smoothing the {\em softmax} layer of the original training data. It then uses the soft labels to train a second neural network that, by way of hidden knowledge transfer, would behave like the first neural network. The key insight is that by training to match the first network, one will hopefully avoid over-fitting against any of the training data. Defensive distillation was later broken by Carlini and Wagner~\cite{CW}.

{\em Adversarial training}~\cite{FGSM}: While effective against the class of adversarial examples a model is trained against, it fails to catch adversarial examples not used during adversarial training. Moreover, robustness comes at a cost of accuracy penalty on clean inputs.

{\em Data transformation approaches} such as compression~\cite{Compression17,Compression18}, augmentation~\cite{Augmentation17}, cropping~\cite{Cropping17}, and randomization~\cite{Rand15,Rand18} have also been proposed to thwart adversarial examples. While these lines of defenses are effective against attacks constructed without the knowledge of the transformation methods, all it takes an attacker to bypass such countermeasures is to employ novel data transformation methods. Like adversarial training, these approaches also succeed at a cost of accuracy loss on clean data.

{\em Gradient masking approaches} (e.g., \cite{Thermo-Encode18,PixelDefend18}) obscure the gradient from a white-box adversary. However, Athalye et al.~\cite{Gradient-Masking18} later broke numerous defenses in this family.

\textbf{Certified Defenses}.
To  obtain a theoretically justified guarantee of robustness, certified defenses have been recently proposed~\cite{wong2018provable,raghunathan2020certified,lecuyer2019certified,RandomSmoothing19}. The key idea is to provide provable/certified robustness of ML models under attack.

Lecuyer et al.\cite{lecuyer2019certified} prove robustness lower bound with differential privacy using Laplacian and Gaussian noise after the first layer of a DNN. Li et al.~\cite{Certified-AdditiveNoise19} build on~\cite{lecuyer2019certified} to prove certifiable lower bound using Renyi divergence. To stabilize the effect of Gaussian noise, Cohen et al.~\cite{RandomSmoothing19} propose randomized smoothing with Gaussian noise to guarantee robustness in $l_2$ norm. They do so by turning any classifier that performs well under Gaussian noise into a new certifiably robust classifier

\textbf{Moving Target Defenses}.
Network and software security has leveraged numerous flavors of MTD including randomization of service ports and address space layout randomization~\cite{MTD-Survey18,MTD-book}. Recent work has explored MTD for defending adversarial examples~\cite{fMTD19,MTDeep19,EI-MTD}.\\
fMTD~\cite{fMTD19} creates fork-models via independent perturbations of the base model and retrains them. Fork-models are updated periodically when the system is in an idle state. Given an input, predicted label is decided by majority vote.\\
MTDeep~\cite{MTDeep19} uses diverse DNN architectures (e.g., CNN, HRNN, MLP) in a manner that reduces transferability between model architectures using a measure called {\em differential immunity}. Through Bayesian Stackelberg game, MTDeep chooses a model to classify an input. Despite diverse model architectures, MTDeep suffers from a small model pool size.\\
EI-MTD~\cite{EI-MTD} leverages Bayesian Stackelberg game for dynamic scheduling of student models to serve prediction queries on resource-constrained edge devices. The student models are generated via differential distillation from an accurate teacher model that resides on the cloud.\\
\textbf{\sysname{} vs. Prior Moving Target Defenses}.
\sysname{} makes several advances beyond the current state-of-the-art. Compared to EI-MTD~\cite{EI-MTD}, \sysname{} avoids inheritance of adversarial training limitations by adversarially training a subset of student models instead of the base model. Unlike EI-MTD that results in lower accuracy on clean data after adversarial training, \sysname{}'s accuracy on clean data after adversarial training is much better since the accuracy penalty is not inherited by student models. Instead of adding a regularization term during training, \sysname{} uses distinct transformed training data to retrain student models and preserve base model accuracy. Instead of Bayesian Stackelberg game, \sysname{} uses the most confident model for prediction guided by OOD detection.
With respect to MTDeep~\cite{MTDeep19}, \sysname{} expands the pool size regardless of the heterogeneity of individual models and uses average transferability rate to estimate attack transferability. On scheduling strategy, instead of Bayesian Stackelberg game \sysname{} uses the most confident model powered by OOD detection-based routing of inputs to most suitable models.
Unlike fMTD~\cite{fMTD19}, \sysname{} goes beyond retraining perturbed fork models and adversarially trains a subset of the model pool to harden the whole pool against adversarial example attacks. In addition, instead of majority vote, \sysname{} picks the most confident model for prediction. For pool renewal, instead of waiting when the system is idle, \sysname{} takes a rather safer and transparent approach and renews an expired pool seamlessly on-the-fly. 

\section{Approach}\label{sec: approach}

\begin{figure*}[t!]
    
    \centering
    \includegraphics[width=.9\textwidth]{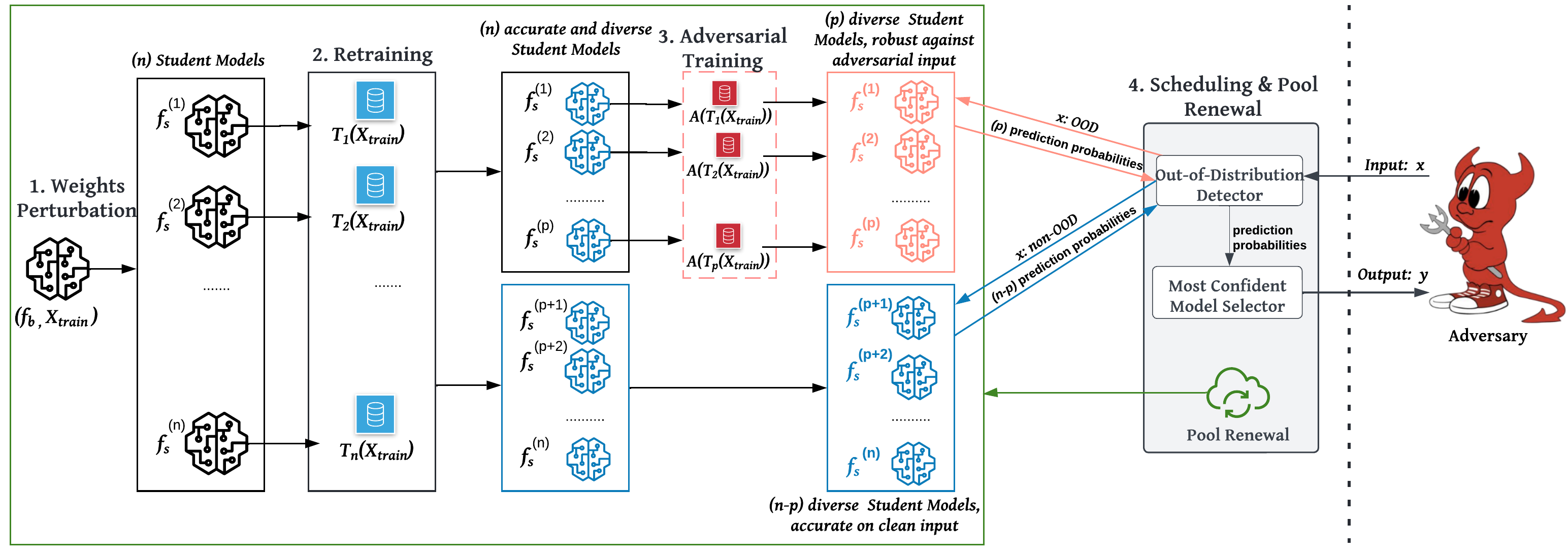}
   \vspace*{-1em}
    \caption{\sysname{} system overview illustrated with model pool generation (1-3) and scheduling and model pool renewal (4).}
    \label{fig:rmtd}
     \vspace*{-2em}

\end{figure*}

We first describe \sysname{} at a high level in Section \ref{subsec:approach-overview} and then dive deeper into details in Sections \ref{subsec:model-gen} - \ref{subsec:renewal}. 
\vspace{-2em}
\subsection{Overview} \label{subsec:approach-overview}
Figure \ref{fig:rmtd} illustrates \sysname{}. The key intuition in \sysname{} is making a model a moving target in the face of adversarial example attacks. To do so, \sysname{} deploys a pool of $n$ models instead of a single target model. We call each model in pool a {\em student model}. The model pool is generated from a base model $f_b$ in a manner that introduces sufficient randomness when \sysname{} responds to prediction queries, without sacrificing accuracy of $f_b$ on clean inputs. In response to each prediction query, \sysname{} selects the {\em most suitable model} to predict the label for the input. The selection of the {\em most suitable model} is governed by a scheduling strategy that relies on the prediction confidence of each model on a given input. The deployed pool of $n$ models automatically expires after a query budget $Q_{max}$ is exhausted. An expired model pool is seamlessly replaced by a new pool of $n$ models generated and queued in advance. As a result, repeated/correlated attacks fail due to the moving target nature of the model. In the following, we use Figure \ref{fig:rmtd} to highlight the core components of \sysname{} focusing on \textit{model pool generation}, {\em scheduling}, and \textit{model pool renewal}.

\textbf{Model Pool Generation.}
The model pool generation method is composed of three main steps (1-3 in Figure \ref{fig:rmtd}). Each step aims to tackle one or more of challenges 1--3. As shown in Figure \ref{fig:rmtd}, a highly accurate $f_b$ initially trained on a training set $X_{train}$ is the foundation from which $n$ models that make up the \sysname{} student model pool are generated. Each student model is initially obtained by slightly perturbing the parameters of $f_b$ (Step-1, \textit{weights perturbation}, in Figure \ref{fig:rmtd}). By introducing different and random perturbations to model parameters, we {\em move} $f_b$'s decision function into $n$ different {\em locations} in the prediction space.

Given the randomness of the weights perturbation, Step-1 is likely to result in inaccurate student models compared to $f_b$. Simply using such inaccurate models as part of the pool penalizes prediction accuracy on clean inputs. As a result, we introduce Step-2 to address {\em Challenge-2} by \textit{retraining} the student models so as to boost their accuracy and bring it close enough to $f_b$'s accuracy. However, since the $n$ models are generated from the same $f_b$, their transferability rate of adversarial examples is usually high. To reduce evasion transferability among student models, we use distinct training sets for each student model (Step-2: {\em retraining} in Figure \ref{fig:rmtd}). For this purpose, we harness data transformation techniques (i.e., image transformations) to produce $n$ distinct training sets. In Section \ref{eval:components}, we empirically explore to what extent this measure reduces the transferability rate and addresses {\em Challenge-3}. Finally, a subset of $p$ student models ($p < n$) is adversarially trained (Step-3: {\em Adversarial training} in Figure \ref{fig:rmtd}) as a reinforcement to the MTD core of \sysname{}, which addresses {\em Challenge-1}. In Section \ref{subsec:model-gen}, we explain the motivations and details of each step.

\textbf{Scheduling and Pool Renewal.}
Instead of randomly selecting a model from the model pool or taking the majority vote of the $n$ models (as in~\cite{fMTD19}), for a given input query \sysname{} returns the prediction of the {\em most confident model}. The motivation to pick the most confident model is twofold. First,  the sufficient diversity among the $n$ models, where a subset of the models is pre-hardened with adversarial training (hence perform much better on adversarial inputs) and the remaining models are trained to perform more confidently on legitimate inputs. Second, the routing of an input to either adversarially trained $p$ models or the remaining $n-p$ models is powered by an OOD detection component (Step 4 in Figure \ref{fig:rmtd}).\\
The model pool is automatically renewed after a defender-set query upper-bound $Q_{max}$ is reached. The choice of $Q_{max}$ requires careful consideration of the model pool size ($n$) and the time it takes to generate a new model pool while \sysname{} is serving prediction queries on an active model pool. In Section \ref{subsec:renewal}, we describe the details of the model pool renewal with respect to $Q_{max}$.

\vspace{-1.5em}
\subsection{Student Model Pool Generation}\label{subsec:model-gen}
Using Algorithms \ref{alg:model_gen} and \ref{alg:model_retrain}, we now describe the details of steps 1--3 with respect to challenges 1--3.

\textbf{Step-1: Model Weights Perturbation.}
Shown in Step-1 of Figure \ref{fig:rmtd}, the first step to transform the fixed model $f_b$ into a moving target is the generation of multiple instances of $f_b$. To effectively serve the MTD purpose, the generated instances of $f_b$ need to fulfill two conditions.  First, they need to be {\em sufficiently diverse} to reduce attack transferability among themselves. Second, they need to preserve the accuracy of $f_b$.

By applying $n$ different small perturbations on the model weights $\theta_b$ of $f_b$, we generate $n$ variations of $f_b$ as $f_s =\{f_s^{(1)}, f_s^{(2)}, ..., f_s^{(n)} \}$, and we call each  $f_s^{(i)}$ a {\em student model}. The $n$ perturbations should be sufficient to produce $n$ diverse models that are different from $f_b$. More precisely, higher perturbations lead to a larger distance between the initial $f_b$ and a student model $f_s^{(i)}$, which additionally contribute to greater movement of the decision function of $f_b$. However, the $n$ perturbations are constrained by the need to preserve the prediction accuracy of $f_b$ for each student model $f_s^{(i)}$. As shown in Algorithm \ref{alg:model_gen} (line 2), we perturb the parameters $\theta_b$ by adding noise sampled from the \textit{Laplace} distribution.

Laplace distribution is defined as $\frac{1}{2\lambda}\exp(-\frac{|\theta_b-\mu|}{\lambda})$ \cite{Laplace}. The center of the post-perturbation weights distribution is the original weights $\theta_b$. We fix the mean value of the added Laplace noise as $\mu = 0$ (line 2). The perturbation bound defined by the Laplace distribution is $exp(\lambda)$, which is a function of the noise scale $\lambda$, also called the exponential decay. Our choice of the Laplace mechanism is motivated by the way the exponential function scales multiplicatively, which simplifies the computation of the multiplicative bound $exp(\lambda)$.

However, there is no exact method to find the maximum noise scale $\lambda_{max}>0$ that guarantees acceptable accuracy of a generated student model. Thus, we approximate $\lambda_{max}$ empirically with respect to the candidate student model by incrementally using higher values of $\lambda>0$ until we obtain a maximum value $\lambda_{max}$ that results in a student model with unacceptable accuracy. Additionally, we explore the impact of increasing $\lambda$ on the overall performance of the prediction framework and the transferability rate of evasion attacks across the $n$ models (Section \ref{eval:components}). We note that the randomness of Laplace noise allows the generation of $n$ different student models using the same noise scale $\lambda$. Furthermore, even in case of a complete disclosure of the fabric of our approach, it is still difficult for an adversary to reproduce the exact pool of $n$ models to use for adversarial example generation, due to the random aspect of the model pool generation approach.

\begin{algorithm}[t!]
\small
\KwResult{$f_s$ : student model}
\SetAlgoLined
\textbf{Input:}\break
$f_b$: base model;\break
$X_{train}$ : training set;\break
$X_{test}$ : testing set;\break
$acc_b \gets Accuracy(f_b, X_{test})$;\break
$T_i$ : data transformation function;\break
$\lambda>0$ : noise scale;\break
$\epsilon > 0$ : used to detect the convergence of model training;\break
$max\_acc\_loss$ : allowed margin of accuracy loss between $f_b$ and $f_s$;\break
$adv\_train$ : boolean variable that indicates whether to train student model on adversarial data;\break
$\Lambda$ : mixture of evasion attacks to use for adversarial training when $adv\_train = TRUE$;\break
\textbf{Step-1:}\tcp{model weights perturbation.}\break
$f_s \gets f_b + Lap(0,\lambda)$;\break
\tcp{$Lap(\mu,\lambda)$ returns an array of noise samples drawn from Laplace distribution $\frac{1}{2\lambda}\exp(-\frac{|\theta_b-\mu|}{\lambda})$}

\textbf{Step-2:}\tcp{retraining on transformed data.}\break
$f_s \gets$ retrain($f_s$, $T_i(X_{train})$, $X_{test}$, $\epsilon$, $Adv=FALSE$);\break
$acc_s \gets $Accuracy($f_s$, $X_{test}$);\break
\While{$acc_b - acc_s$ $>$ $max\_acc\_loss$}
{ repeat Step-1 and Step-2 with smaller $\lambda$;}
\textbf{Step-3:}\tcp{retraining on adversarial data.}\break
\If{$adv\_train$}{$f_s \gets $retrain($f_s$, $\Lambda(T_i(X_{train}))$, $X_{test}$, $\epsilon$, $Adv=TRUE$);\break
\tcp{check accuracy loss on clean test set.}\break
$acc_s \gets$ Accuracy($f_s$, $X_{test}$);\break
\While{$acc_b - acc_s > max\_acc\_loss$}
{$f_s \gets$ retrain($f_s$, $T_i(X_{train})$, $X_{test}$, $\epsilon$, $Adv=FALSE$);\break}}

\caption{Student model generation.}
\label{alg:model_gen}
\end{algorithm}

\textbf{Step-2: Retraining on Transformed Data.}
Minor distortions of the parameters of $f_b$ have the potential to reduce the prediction accuracy of the resultant student model. Consequently, Step-1 is likely to produce student models that are less accurate than $f_b$. An accuracy recovery measure is necessary to ensure that each student model has acceptable accuracy close enough to $f_b$. To that end, we retrain the $n$ newly created student models (line 3).

Diversity of the model pool is crucial for \sysname{}'s MTD core such that adversarial examples are less transferable across models ({\em Challenge-3}). In this regard, retraining all student models on $X_{train}$ used for $f_b$ results in models that are too similar to the decision function of $f_b$. It is, therefore, reasonable to use a distinct training set for each student model. To tackle data scarcity, we harness data augmentation techniques to perform $n$ distinct transformations $\{T_1,...,T_n\}$ on the original training set $X_{train}$ (e.g., translation, rotation, etc). The translation distance or the rotation degree are randomly chosen with respect to the validity constraint of the transformed set $T_i(X_{train})$. A transformed sample is valid only if it is still recognized by its original label. In our case, for each dataset, we use benchmark transformations proposed and validated by previous work \cite{Tian2018DetectingAE}. We additionally double-check the validity of the transformed data by verifying whether each sample is correctly predicted by $f_b$.

\begin{algorithm}[t!]
\small
\KwResult{$f_s^{(i)}$ : student model}
\SetKwFunction{Fretrain}{Retrain}
\SetAlgoLined
\SetKwProg{Fn}{Def}{:}{}
  \Fn{\Fretrain{$f_s^{(i)}$, $X_{retrain}$, $X_{test}$, $\epsilon$, $Adv$}}{
  \tcp{For adversarial training we use adversarial test examples for validation.}\break
  \If{$Adv=TRUE$}{$X_{test}\gets\Lambda(X_{test})$}
  $acc_{tmp} \gets$ Accuracy($f_s^{(i)}$, $X_{test}$);\break
  $epochs \gets 0$;\break
  \While{TRUE}{
  $f_s^{(i)}$.train($X_{retrain}$, $epoch=1$);\break
  $acc \gets$ Accuracy($f_s^{(i)}$, $X_{test}$);\break
  \tcp{check training convergence.}\break
  \If{$epochs$ mod($5$) $=0$}{
    \eIf{$|acc - acc_{tmp}|<\epsilon$}{break;\break}
    {$acc_{tmp}\gets acc$;}
    }\break
    $epochs \gets epochs + 1$;\break
    }
   }

\caption{Student model retraining.}
\label{alg:model_retrain}
\end{algorithm}
Algorithm \ref{alg:model_retrain} illustrates the student model retraining function called \texttt{Retrain($f_s$,$X_{retrain}$,$X_{test}$,$\epsilon$,$Adv$)}. It takes as inputs: a student model $f_s$, a retraining set $X_{retrain} = T_i(X_{train})$, a testing set $X_{test}$, a small positive infinitesimal quantity $\epsilon \rightarrow 0$ used for training convergence detection, and a boolean flag $Adv$. As indicated on line 15 of Algorithm \ref{alg:model_gen}, $Adv$ is $FALSE$ since the retraining data does not include adversarial examples. The algorithm regularly checks for training convergence after a number of (e.g., 5) epochs. The retraining convergence is met when the current accuracy improvement is lower than $\epsilon$ (lines 7--13 in Algorithm \ref{alg:model_gen}).

The validity of the selected value of the noise scale $\lambda$ used in Step-1 is decided by the outcome of Step-2 in regaining the prediction accuracy of a student model. More precisely, if retraining the student model (i.e., Step-2) does not improve the accuracy, then the optimisation algorithm that minimizes the model's loss function is stuck in a local minimum due to the significant distortion brought by weight perturbations performed in Step-1. In this case, we repeat Step-1 using a lower $\lambda$, followed by Step-2. The loop breaks when the retraining succeeds to regain the student model's accuracy, which indicates that the selected $\lambda$ is within the maximum bound $\lambda < \lambda_{max}$ (Algorithm \ref{alg:model_gen}, lines 4-6). For more control over the accuracy of deployed models, we define a hyperparameter $max\_acc\_loss$, that is configurable by the defender. It represents the maximum prediction accuracy loss tolerated by the defender.
\\

\textbf{Step-3: Adversarial Training a Subset of $p$ Student Models.} 

To motivate the need for Step-3, let us assess our design using just Step-1 and Step-2 with respect to challenges 1--3.
Suppose \sysname{} is deployed based only on Step-1 and Step-2, and for each input it picks the most confident student model and returns its prediction. On clean inputs, the MTD strategies introduced in Step-1 (via model weights perturbation) and Step-2 (via retraining on transformed training data) make \sysname{} a moving target with nearly no loss on prediction accuracy. On adversarial inputs, an input that evades student model $f^{(i)}_s$ is less likely to also evade another student model $f^{(j)}_s$ because of the significant reduction of transferability between student model predictions because of Step-2. However, due to the exclusive usage of clean inputs in Step-1 and Step-2, an adversarial example may still fool a student model on first attempt. We note that the success rate of a repeated evasion attack is low because the randomness introduced in Step-1 disarms the adversary of a stable fixed target model that returns the same prediction for repeated queries on a given adversarial input. To significantly reduce the success of one-step attacks, we introduce selective adversarial training to reinforce the MTD strategy built through Step-1 and Step-2. More precisely, we perform adversarial training on a subset $p < n$ models from the $n$ student models obtained after Step-2 (lines 7--13 in Algorithm \ref{alg:model_gen}). We note our choice of adversarial training is based on the current state-of-the-art defense. In principle, a defender is free to use a different (possibly better) method than adversarial training. 

\textbf{Why adversarial training on $p < n$ student models?} 
There are three intuitive alternatives to integrate adversarial training to the MTD strategy: $(a)$ adversarial training of $f_b$ before Step-1; $(b)$ adversarial training  of all $n$ student models after Step-2; or $(c)$ adversarial training of a subset of student models after Step-2. As noted by prior work \cite{EI-MTD}, $(a)$ is bound to result in an inherited robustness for each student model, which costs less execution time compared to adversarial training of $n$ student models. However, in this case, the inherent limitation of adversarial training, i.e., accuracy reduction on clean inputs, is also inherited by the $n$ student models. Alternative $(b)$ suffers from similar drawbacks. By adversarially training all $n$ models, while making individual student models resilient against adversarial examples, we risk accuracy loss on clean data. Considering the drawbacks of $(a)$ and $(b)$, we pursue $(c)$. In particular, we select $p < n$ student models for adversarial training (lines 7--13 in Algorithm \ref{alg:model_gen}). Consequently, the remaining $n-p$ models remain as accurate as $f_b$ on clean data in addition to being diverse enough to reduce attack transferability among them. 

\textbf{Adversarial training approach.} 
We now explain the details of adversarially training a student model with respect to Algorithm \ref{alg:model_gen}.
Once again, the \texttt{Retrain} function illustrated in Algorithm \ref{alg:model_retrain} is invoked using different inputs (line 11). For instance, $X_{retrain}=\Lambda(T_i(X_{train})$, which indicates that $f_s$ is trained on adversarial examples, is generated by performing a mixture of evasion attacks $\Lambda$ on the transformed training set $T_i(X_{train})$ specified for student model $f_{s}^{(i)}$. To reduce accuracy decline on clean data, we shuffle the training set with additional clean samples from $T_i(X_{train})$. Furthermore, we use more than one evasion attack with different perturbation bounds $||\delta||<\xi$ for adversarial samples generation to boost the robustness of the student model against different attacks (e.g., C\&W~\cite{CW}, gradient-based~\cite{FGSM}, etc). Like Step-2, the training convergence is reached if the improvement of the model's accuracy on adversarial examples (i.e., the robustness) recorded periodically, i.e., after a number (e.g., 5) of epochs, becomes infinitesimal (lines 7--13 in Algorithm \ref{alg:model_retrain}).

\textbf{How to choose the values of $p$ and $n$?} The values $p$ and $n$ are defender-chosen hyper-parameters. Ideally, larger $n$ favors the defender by creating a wider space of movement for a model's decision function (thus creating more uncertainty for repeated or correlated attacks). However, in practice $n$ is conditioned by the computational resources available to the defender. Therefore, here we choose not to impose any specific values of $n$.  We, however, recall that $Q_{max}$ is proportional with the time needed to generate a pool of $n$ models. Therefore, it is plausible that $n$ need not be too large to lead to a long extension of the expiration time of the pool of $n$ models due to a longer period $T_n$ needed to generate $n$ models that causes a large value of $Q_{max}$. Regarding the number of adversarially-trained models ($p$), there is no exact method to select an optimal value. Thus, we empirically examine all possible values of $p$ between $0$ and $n$ to explore the performance change of \sysname{} on clean inputs and adversarial examples. Additionally, we explore the impact of the value $p$ on the transferability rate across the $n$ student models (results are discussed in Section \ref{eval:components}).

\vspace{-1em}
\subsection{Scheduling Strategy}\label{subsec: scheduling}
In the following, what we mean by {\em scheduling strategy} is the act of selecting the model that returns the label for a given input. There are multiple alternatives to reason about the scheduling strategy. Randomly selecting a model or taking the majority vote of all student models are both intuitive avenues. However, random selection does not guarantee effective model selection and majority vote does not consider the potential inaccuracy of adversarially trained models on clean queries.

\textbf{Most Confident Model.} We rather adopt a strategy that relies on the confidence of each student model. More precisely, given an input $x$, \sysname{} first queries each student model and returns the prediction of the most confident model. Given a query $x$, the scheduling strategy is formally defined as: $\argmax \{f_s^{(1)}(x),...,f_s^{(n)}(x)\}$.

\textbf{OOD-Powered Scheduling.}
On top of \sysold{}, we extend our scheduling strategy with a pre-cursor decision component that determines whether an input needs to be predicted by the most confident model from the pool of $p$ adversarially trained models or the most confident model of the remaining $n-p$ undefended models. Cognizant of the cost of robustness using adversarial training (i.e., accuracy loss on benign data), we aim to guide the scheduling approach to send benign queries to the $n-p$ highly accurate undefended models, while any adversarial query is sent to the remaining $p$ models that are defended by adversarial training. To this end, as motivated in section \ref{sec: intro}, we leverage recent advances in OOD detection to separate between benign in-distribution queries from potential adversarial examples that are most likely OOD. Specifically, we adopt the current state-of-the-art method called \textit{SSD} that trains a {\em self-supervised outlier detector} through learning a feature representation of the data distribution $\mathbb{P}_{train}$ used to train the target model $f$. Given an input sample $x$, SSD computes how far $x$ is distant from the training data distribution using the Mahalabolis distance metric ($M(x,\mathbb{P}_{train})$) ~\cite{SSD} (Section \ref{SSD}). Although, most adversarial examples are known to be OOD examples~\cite{Amich_2021}, in order to effectively use SSD as an adversarial examples detector, a threshold definition ($\tau$) is necessary to separate between potential OOD adversarial examples and benign in-distribution queries that exhibit tolerable distribution shifts (i.e., $M(x,\mathbb{P}_{train})<\tau$) from those that are far away from $\mathbb{P}_{train}$ due to potential adversarial perturbations (i.e., $M(x,\mathbb{P}_{train}) \ge \tau$).

\begin{figure}[htp]
\vspace{-1em}
    \centering
    \includegraphics[height=.75\columnwidth]{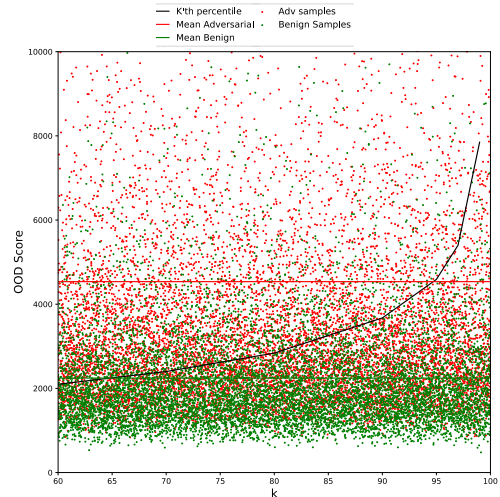}
    \vspace{-1em}
    \caption{$k^{th}$ percentile threshold for OOD detection.}
    \label{fig:galaxy}
    
    \vspace{-2em}
\end{figure}

\textbf{OOD Detection Threshold Determination.}
The OOD distance score of an adversarial example $M(x+\delta,\mathbb{P}_{train})$ tends to be higher than the distance of benign inputs, i.e., $M(x,\mathbb{P}_{train})$, due to distribution shifts that can be caused by adversarial perturbations. In Figure \ref{fig:galaxy}, we compare the OOD scores (y-axis) of FGSM samples on CIFAR10 (red points) to the OOD scores of benign samples (green points). Figure \ref{fig:galaxy} visually confirms that most of the adversarial samples (red) have higher OOD scores compared to the benign samples (green). 
However, a threshold is needed to define the maximum allowable distance $M(x,\mathbb{P}_{train})$ that a sample $x$ can record in order to be considered benign. To this end, we refer to a subset of benign in-distribution training data that we use for threshold selection. We call it tuning data $X_{\tau} \sim \mathbb{P}_{train}$. Intuitively, we can select as a threshold the maximum distance score recorded by samples in the tuning data $X_\tau$. Formally,
    $ \tau = \max_{x \in X_\tau} M(x,\mathbb{P}_{train})$.
With respect to Figure \ref{fig:galaxy}, such a threshold would fail in practice. Particularly, we observe few green points (benign) that have OOD scores similar or higher than most red points (adversarial). Such unexpected outliers in $X_\tau$ can make $\tau$ very high. Therefore, it would classify most of adversarial examples as benign. Furthermore, the threshold should be agnostic to the studied attack (e.g., FGSM) and dataset (e.g., CIFAR10). Hence, a fixed threshold is not a suitable choice. Moving forward, in Figure \ref{fig:galaxy}, we explore the $k^{th}$ \textit{percentile} (black curve) of $X_\tau$ as a threshold with respect to different values of $k$ (x-axis). In other words, we select as threshold $\tau$ the lowest OOD score in $X_\tau$ that is greater than a $k \%$ of all scores recorded by $x \in X_\tau$. Formally,
\begin{equation}
     \tau = k^{th} Percentile\{ M(X_\tau),\mathbb{P}_{train})\}
 \end{equation}
 
$k$ is a variable that can be changed with respect to the studied dataset and attacks. For instance, in Figure \ref{fig:galaxy}, $k$ in the range $[85^{th}, 95^{th}]$ is above the mean of benign samples ($X_\tau$) and below the mean of their adversarial counterparts ($X'_\tau$). Consequently, one can separate between benign and adversarial samples while ignoring outliers of benign samples that have an OOD score higher than $k\%$ of all other samples in $X_\tau$. By leveraging the $k^{th} percentile$ selection strategy we find a tradeoff between accurately detecting adversarial examples and avoiding false positives. In section \ref{eval:robust}, we investigate the effectiveness of the scheduling strategy compared to the most confident model approach.

\subsection{Model Pool Renewal} \label{subsec:renewal}
Given that $n$ finite, with enough time, an adversary can ultimately build knowledge about the prediction framework through a series of queries. For instance, if the adversary correctly guesses model pool size $n$, it is possible to map which model is being selected for each query by closely monitoring the prediction patterns of multiple examples. Once compromised, the whole framework becomes a sitting target since its movement is limited by the finite number $n$.

On way to avoid such exposure is abstaining from responding to a ``suspicious" user \cite{goodfellow2019research}. However, given the difficulty of precisely deciding whether a user is suspect based solely on the track of their queries, this approach has the potential to result in a high abstention rate, which unnecessarily leads to denial of service for legitimate users. We, therefore, propose a relatively expensive yet effective method to ensure the continuous mobility of the target model without sacrificing the quality of service. More precisely, we actively update the pool of $n$ models when a query budget upper bound $Q_{max}$ is reached. To maintain the quality of service in terms of query response time, the update needs to be seamless. To enable seamless model pool update, we ensure that a buffer of batches of $n$ student models is continuously generated and maintained on stand-by mode for subsequent deployments.

The choice of $Q_{max}$ determines how dynamic the target model under the condition:
\textit{"the buffer of pools of models is never empty at a time of model batch renewal"}.\\ 
Suppose at time $t$ the buffer contains $K_t$ pools of models. A new pool is removed from the buffer after every period of $Q_{max}$ queries and a clean-slate pool is activated. Thus, the buffer is exhausted after $K_t.Q_{max}$ number of queries. Supposing that the per-query response time is $T_q$ and the generation of a pool of $n$ models lasts a period of $T_n$, the above condition is formally expressed as:
$\begin{array}{rrclcl}
\displaystyle
K_t.Q_{max}.T_q > T_n \textrm{, s.t.} & K_t > 0.
\end{array}$
The inequality implies that the time to exhaust the whole buffer of pools must be always greater than the duration of creating a new pool of $n$ models. Additionally, it shows that $Q_{max}$ is variable with respect to the time $t$ and the number of models in one pool $n$. Ideally, $Q_{max}$ should be as low as possible to increase the mobility rate of target model. Thus, the optimal solution is $\ceil {\frac{T_n}{K_t.T_q} }$. 
\vspace{-1.5em}

\section{Evaluation}\label{sec: eval}
We now evaluate \sysname{}.
Section \ref{subsec:setup} presents experimental setup. Section \ref{eval:robust} compares \sysname{} with the undefended base model, adversarially trained base model, and \sysold. Section \ref{scheduling} evaluates the effectiveness of the scheduling strategies. Section \ref{mtd_impact} examines the impact of dynamic scheduling and model pool renewal. Finally, Section \ref{eval:components} sheds light on the impact of individual \sysname{} components on robustness and attack transferability.
\vspace{-1em}
\subsection{Experimental Setup}
\label{subsec:setup}
To enable a fair comparison with \sysold{}, we evaluate \sysname{} using the same experimental setup introduced in \sysold{}'s paper \cite{Amich_2021}. We also experiment both approaches against additional attacks.

\textbf{Datasets:}
We use two benchmark datasets: MNIST~\cite{MNIST} and CIFAR10~\cite{cifar}. We use $5$K test samples of each dataset to perform $5$K queries for evaluation.



\textbf{Attacks:}
We use three \textit{white-box} attacks (FGSM~\cite{FGSM}, PGD \cite{PGSM}, and C\&W~\cite{CW}) and one \textit{black-box} attack (SPSA \cite{uesato2018adversarial}). Details of these attacks appear in \sysold{} \cite{Amich_2021}. For C\&W, PGD, and FGSM, we assume the adversary has white-box access to $f_b$. For SPSA, the adversary has oracle access to \sysname{}. We carefully chose each attack to assess our contribution claims. For instance, C\&W, one of the most effective white-box attacks, is suggested as a benchmark for ML robustness evaluation~\cite{carlini2019evaluating}. PGD is a widely used gradient-based attack that can reach a higher evasion rate, while FGSM is fast and scalable on large datasets and generalizes across models ~\cite{kurakin2017adversarial}. In addition, the relatively high transferability rate of FGSM attacks across models makes it suitable to evaluate the effectiveness of \sysname{}'s different components to reduce attack transferability across student models. To explore \sysname{}'s robustness against query-based black-box attacks, we employ SPSA since it performs multiple correlated queries to craft adversarial examples. For all attacks we use a perturbation bound $||\xi||_\infty < 0.3$.

\textbf{Base Models:}
As base models we reuse the same models previously introduced in \cite{Amich_2021}. Notably, 6-layer CNN model that reaches a test accuracy of $99.72\%$  (i.e.,``MNIST-CNN") and a CNN CIFAR10 model \cite{CopyCat18} that reaches an accuracy of $83.63\%$ on a test set of 5K (i.e., ``CIFAR10-CNN").

\textbf{Baseline Defenses:}
In accordance with the baseline defenses adopted in \cite{Amich_2021}, for both datasets, we compare \sysname{} with an \textit{undefended fixed} model, an \textit{adversarially-trained fixed} model, and \sysold{}. 

\textbf{Hyper-parameters:}
We refer to the same hyper-parameters in \cite{Amich_2021}. Notably, we use $5$ pools of models where each of size $n=10$, $Q_{max}=1$K.  

In Table \ref{tab:evasion_res}, for MNIST, we fix $\lambda=0.1$, $p=5$ and for CIFAR10, we use  $\lambda=0.05$ and $p=8$ or $p=9$. More details about hyper-parameters tuning is explained in \cite{Amich_2021}.

\textbf{Metrics:}
As adopted in \cite{Amich_2021}, we use as evaluation metrics,
\textit{\underline{Accuracy}}
and \textit{\underline{Average Transferability Rate (ATR)}} (\cite{Amich_2021}).
We compute {\em ATR} across all $n$ models to evaluate the effectiveness of the data transformation measures at reducing attack transferability. To compute the transferability rate from model $f^{(i)}_s$ to another model $f^{(j)}_s$, we calculate the rate of adversarial examples that succeeded on $f^{(i)}_s$ that also succeed on $f^{(j)}_s$. Across all student models, {\em ATR} is computed as:


$\displaystyle 
ATR = \frac{1}{n (n-1)}\sum_{i=1}^{n} \sum_{\substack{j=1 \\ j\ne i}}^{n} \frac{N_{adv}(f^{(i)}_s \rightarrow f^{(j)}_s)}{N_{adv}(f^{(i)}_s)},$\\

where $N_{adv}(f^{(i)}_s) = \{x' \in X'_{test}; f^{(i)}_s(x') \ne y_{true}\}$ is the number of adversarial examples that fooled $f^{(i)}_s$ and $N_{adv}(f^{(i)}_s \rightarrow f^{(j)}_s) = \{x' \in N_{adv}(f^{(i)}_s); f^{(j)}_s(x') \ne y_{true}\}$ is the number of adversarial examples that fooled $f^{(i)}_s$ that also fool $f^{(j)}_s$.

\begin{table*}[t!]
\centering
  \scalebox{0.80}{
   \begin{tabular}{|c!{\color{black}\vrule}c!{\color{black}\vrule}c!{\color{black}\vrule}!{\color{black}\vrule}!{\color{black}\vrule}c!{\color{black}\vrule}c!{\color{black}\vrule}c!{\color{black}\vrule}c!{\color{black}\vrule}c!{\color{black}\vrule}c!{\color{black}\vrule}c!{\color{black}\vrule}c!{\color{black}\vrule}c!{\color{black}\vrule}}
       \hline
       
& \multicolumn{5}{c!{\color{black}\vrule}}{\textbf{MNIST-CNN Accuracy}} & \multicolumn{5}{c!{\color{black}\vrule}}{\textbf{CIFAR10-CNN Accuracy}}\\
\hline
\multirow{2}{*}{\textbf{Attack}}& \multirow{2}{*}{\textbf{Undefended}} & \multirow{2}{*}{\textbf{\thead{Adv-Train} }} & \multirow{2}{*}{\textbf{\sysold}} &
\multicolumn{2}{c!{\color{black}\vrule}}{\thead{\textbf{\sysname{}}}}&\multirow{2}{*}{\textbf{Undefended}} &  \multirow{2}{*}{\textbf{\thead{Adv-Train}}} & \multirow{2}{*}{\thead{\textbf{\sysold{}} ($p=8$, $p=9$)}} & \multicolumn{2}{c!{\color{black}\vrule}}{\thead{\textbf{\sysname{}} ($p=9$)}}\\ 
       \cline{5-6}\cline{10-11}
       &&&&$\tau =95^{th}$&$\tau =90^{th}$&&&&$\tau = 95^{th}$&$\tau =90^{th}$\\
       \hline
       No Attack & 99.72\%& 97.17\% &99.04\%&\textbf{99.58\%}&\textbf{99.58\%}& 83.63\%& 75.37\% &\textbf{84.64\%}, 82.65\%&\textbf{83.34\%}&79.44\% \\ 
       \hline\hline
       FGSM~\cite{FGSM} & 9.98\%&  42.38\% &71.43\%&\textbf{86.48\%}&\textbf{86.48\%}&19.98\%&36.62\%&36.44\%, 38.78\%&\textbf{46.82\%}&\textbf{46.82\%}\\ 
       \hline
       PGD~\cite{PGSM} & 0.3\%&  4.14\% &58.02&\textbf{94.08\%}&\textbf{94.08\%}&10.13\%&14.47\%&10.14\%, 10.28\%&\textbf{28.19\%}&\textbf{52.04\%}\\
       \hline
       C\&W~\cite{CW} & 0.0\%& 0.0\% &\textbf{97.75\%}&92.28\%&96.41\%&1.25\%&1.34\%&44.50\%, 40.91\%&\textbf{45.08\%}&\textbf{44.67\%}\\ 
       \hline
       SPSA~\cite{uesato2018adversarial} &29.04\% & 59.43\%&97.77\%&\textbf{98.07}&\textbf{98.62\%}&38.96\%&59.08\%&60.85\%, 62.83\%&\textbf{70.06\%}&\textbf{75.13\%} \\ 
       \hline
   \end{tabular}}
 
  \vspace*{0.25em}
\caption{\sysname{} robustness compared to an undefended, adversarially trained fixed model, and \sysold.}

\label{tab:evasion_res}
\vspace{-3.5em}
\end{table*}
 \vspace{-1em}
\subsection{Robustness Against Evasion Attacks}
\label{eval:robust}
Based on results summarized in Table \ref{tab:evasion_res}, we now evaluate the robustness of \sysname{} compared with the undefended base model, adversarially trained base model, and \sysold{} across the 4 reference attacks. Note that we are particularly interested in the difference in robustness between \sysold{} (scheduling: based on most confident model) and \sysname{} (scheduling: powered by OOD detector).

\textbf{Robustness in a nutshell:} Across all attacks and threat models, both \sysold{} and \sysname{} are more robust than \textit{adversarial training} for both datasets. On MNIST, across all four attacks, \sysold{} and \sysname{} significantly outperform adversarial training by an average of $\approx 55\%$ and $\approx 67\%$, respectively. On CIFAR10, Table \ref{tab:evasion_res} suggests similar results. On CIFAR10, \sysold{} improves robustness by $\approx 2\%$ on FGSM and $\approx 4\%$ on SPSA when $p=9$. With regards to C\&W, it drastically improves robustness compared to the baseline models (i.e., $\approx 41\%$) while we observe a small decrease against PGD (more explanation in \ref{scheduling}). Additionally, \sysname{}, further improves our results. It outperforms adversarial training across all attacks by an average of $22\%$ for both OOD detection threshold configurations. These findings show that \sysname{} is more robust than \sysold{} when powered by OOD detection for input scheduling.

\textbf{Accuracy loss on clean data:} Table \ref{tab:evasion_res} indicates that, unlike adversarial training on a fixed model, \sysname{} does not sacrifice accuracy to improve robustness. For instance, while adversarial training drops the accuracy of the undefended MNIST-CNN by $\approx 3\%$, both \sysold{} and \sysname{} maintain it close to its original value ($>99\%$). Similar results are observed on CIFAR10. On one hand, even after using $9$ adversarially-trained models ($p=9$) of $n=10$ student models for each pool, the accuracy loss is $\leq1\%$ for \sysold{}. As for \sysname{}, it becomes marginal when the threshold is set to the $95^{th}$ percentile ($\tau = 95^{th}$). On another hand, adversarial training sacrifices $8.26\%$ of the original accuracy of CIFAR10-CNN. These results are linked to the effectiveness of the adopted scheduling strategy to assign clean queries mostly to student models that are not adversarially trained, thus, accurate on clean data (more details in \ref{scheduling}). 
Furthermore, \sysname{} can improve the original accuracy for lower values of $p$. For instance, Table \ref{tab:evasion_res} shows an improvement of $1\%$ in the accuracy on CIFAR10 clean data when using \sysold{} with $p=8$, compared to the undefended baseline model.
In conclusion, our findings indicate that \textit{\sysname{} is much more robust compared to adversarial training on fixed model and \sysold{} while preserving the original accuracy of the undefended $f_b$}.

\textbf{Robustness against C\&W:} Inline with the state-of-the-art, Table \ref{tab:evasion_res} shows that C\&W is highly effective on the baseline fixed models. For both datasets, even after adversarial training, C\&W attack can maintain its high attack accuracy for both datasets ($100\%$ on MNIST and $\approx 99\%$ on CIFAR10). However, it significantly fails to achieve the same attack accuracy on \sysname{}. For instance, \sysold{} increases the robustness against C\&W data by $\approx 97\%$ for MNIST and $\approx40\%$ for CIFAR10 compared to adversarial training on fixed model. This significant improvement in robustness is brought by the moving target aspect included in \sysname{}. More precisely, given the low transferability rate of C\&W examples across different models, C\&W queries that are easily effective on $f_b$ are not highly transferable to the student models, hence less effective on \sysname{}. More discussion that reinforces this insight is presented in Section \ref{eval:components}.

\textbf{Robustness against FGSM and PGD:} Although known to be a less effective attack than C\&W on the fixed models, FGSM performs better on \sysname. For instance, \sysold{} accuracy on FGSM data is $26.32\%$ less for MNIST and $\approx 3\%$ less for CIFAR10, compared to C\&W. These findings are explained by the high transferability rate of FGSM examples across student models and $f_b$. Similar behavior is observed for PGD. However, despite the transferability issue, \sysold{} still improves robustness on FGSM and PGD data. More discussion about the transferability effect on \sysname{} are provided in Section \ref{eval:components}. It is noteworthy that, once again, \sysname{} further improves the robustness against FGSM and PGD by an average margin of $\approx 25\%$ on MNIST and $\approx 20\%$ on CIFAR10 (averaged across FGSM and PGD). These results confirm the importance of introducing the OOD detector to separate between benign queries and potential adversarial queries.

\textbf{Robustness against SPSA:} We now turn to a case where the adversary has a black-box prediction API access to \sysname{} to issue multiple queries. The adversary performs iterative perturbations of an input to reach the evasion goal. Inline with the adversary's goal here, SPSA performs multiple queries using different variations of the same input to reduce the SPSA loss function. Table \ref{tab:evasion_res} shows that \sysname{} is more robust on SPSA than the two baseline fixed models. Due to its dynamic characteristic, \sysname{} is a moving target. Hence, it can derail the iterative query-based optimization performed by SPSA. More analysis into the impact of the dynamic aspect is discussed in Section \ref{mtd_impact}. Consistent with results against white-box attacks, \sysname{} is also better than \sysold{} against the studied black-box attack, SPSA. In Section, \ref{scheduling}, analyze scheduling history of \sysold{} vs \sysname{} over benign queries and adversarial queries of different attacks to verify highlight the improved robustness by \sysname{} over \sysold{}.


\textbf{Robustness across datasets:}
\sysname{} performs much better on MNIST than CIFAR10. This observation is explained by various factors. First, CNN models are highly accurate on MNIST ($>99\%$) than CIFAR10 ($\approx 84\%$). Second, across all attacks, on average, \textit{adversarial training} is more effective on MNIST; it not only leads to higher robustness (i.e., $\approx +31\%$ for MNIST compared to $\approx +19\%$ for CIFAR10) it also sacrifices less accuracy on clean MNIST data (i.e.,  $-2.55\%$ for MNIST compared to $-8.26\%$ for CIFAR10). Consequently, the $p$ adversarially-trained student models on MNIST are more robust and accurate than the ones created for CIFAR10. Finally, CIFAR10 adversarial examples are more transferable across student models than MNIST. More results about the transferability factor are detailed in Section \ref{eval:components}.

\noindent \fbox{\parbox{.96\linewidth}{
{\small
 \textbf{Observation 1:} Compared to adversarial training, both \sysold{} and \sysname{} improve robustness to adversarial examples on both benchmark datasets (MNIST, CIFAR10) for both white-box and black-box attacks. This is achieved without sacrificing accuracy on clean data}.}}
 
\subsection{Effectiveness of Scheduling Strategy}
\label{scheduling}
\begin{figure*}[t!]
    
    \centering   
    \includegraphics[width=.96\textwidth]{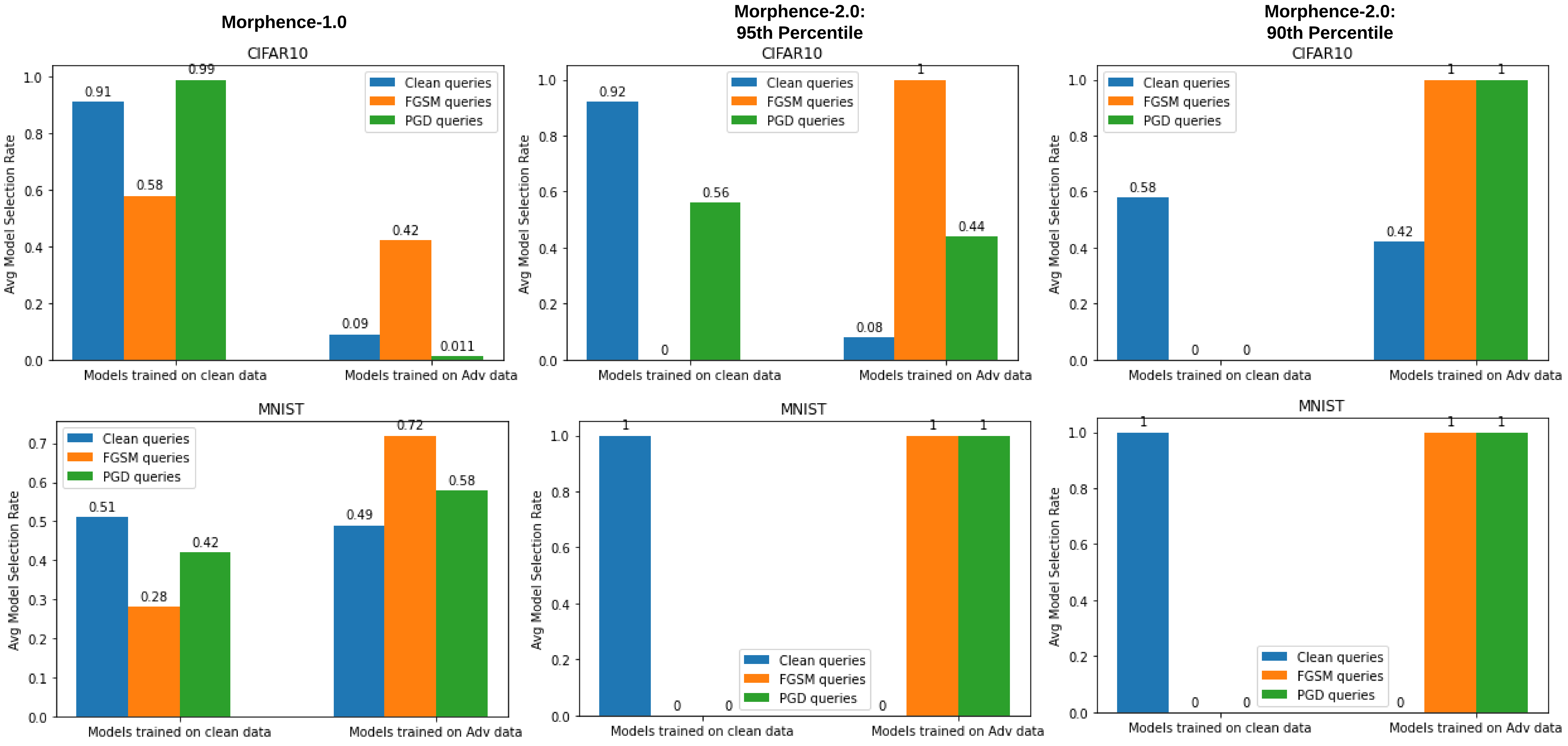}
    \vspace{-1em}
    \caption{Scheduling results of \sysold{} compared to \sysname{}.}
    \label{fig:scheduling}
\vspace{-2em}
\end{figure*}

As illustrated in Section \ref{subsec: scheduling}, given an input query, a scheduling strategy selects the most suitable model for the prediction task. We recall that in \sysold{} the scheduler simply picks the most confident model from the active batch of all the $n$ models regardless of the nature of the input, while in \sysname{} we extend the scheduling approach with another decision layer that separates OOD adversarial inputs from in-distribution benign ones. Our approach aims to precisely assign adversarial examples to the most confident model from the $p$ adversarially trained models, while attributing benign queries to the remaining $n-p$ undefended models that are more accurate on benign data (explained in section \ref{sec: approach}). In Figure \ref{fig:scheduling}, we keep track of the scheduling history of different \sysold{} and \sysname{} with respect to clean (benign) queries and adversarial queries. In the following, as we interpret the scheduling history recorded by each design on both datasets, we refer to Table \ref{tab:evasion_res} to explain the impact of the scheduling precision on improving the robustness of our approach.

On CIFAR10, \sysold{} successfully assigned $91\%$ of the clean queries (blue bar) to undefended models, which explains its high accuracy on benign data compared to adversarial training that sacrifices $\approx 8\%$ of the original accuracy (Table \ref{tab:evasion_res}). However, it fails to correctly handle adversarial queries. Almost all PGD queries (green bar) are falsely assigned to undefended models which explain the very low accuracy of \sysold{} against PGD attack recorded in Table \ref{tab:evasion_res}. As for the FGSM queries (orange bar), \sysold{} exhibits a better scheduling precision, however, it assigns only $42\%$ of FGSM queries to the adversarially-trained models. On the contrary, powered by the OOD detection, \sysname{} succeeds to assign all FGSM queries to the adversarially-trained models which reflects the improvement in robustness observed in Table \ref{tab:evasion_res}. Same results are observed for the PGD queries when $k=90^{th}$ is used as threshold. We recall that using a lower threshold $\tau$ favors more the detection of adversarial examples while it tolerates some false positives on benign data. This tradeoff is observed on CIFAR10 results. More precisely, while \sysname{} with $\tau = 95^{th}$ leads to better scheduling precision on benign data ($92\%$) it scores only $44\%$ on PGD data. A reduction of the threshold to $\tau = 90^{th}$ leads to $100\%$ precision on $PGD$ queries while it scores worse on clean data. These findings explain why, in Table \ref{tab:evasion_res}, \sysname{} with $\tau=90^{th}$ records better robustness but it sacrifices more accuracy on CIFAR10 benign data. 

On MNIST, \sysname{} scores a perfect scheduling precision for both threshold configurations. In other words, all adversarial queries are successfully fed to adversarially trained models, while all clean queries are assigned to undefended models. Particularly, Figure \ref{fig:scheduling} shows that, for both threshold configurations, unlike \sysold{}, \sysname{} assigns $100\%$ of clean queries only to undefended models, while it attributes $100\%$ of FGSM and PGD queries to adversarially trained models. Such high scheduling precision on MNIST explains the significant improvement in robustness by \sysname{} in Table \ref{tab:evasion_res}.

\noindent \fbox{\parbox{.96\linewidth}{
{\small
\textbf{Observation 2:} 
 The OOD-powered scheduling in \sysname{} is more effective than the scheduling strategy in \sysold{} which relies on the most confident model.\\
 \textbf{Observation 3:}  In \sysname{}, a careful configuration of the percentile order $k$ is mandatory to find the defender-desired trade-off between adversarial example detection (higher robustness) and benign example detection (high accuracy on benign data)}.}}

\begin{figure}[t!]
    
    \centering
    \includegraphics[scale=.55]{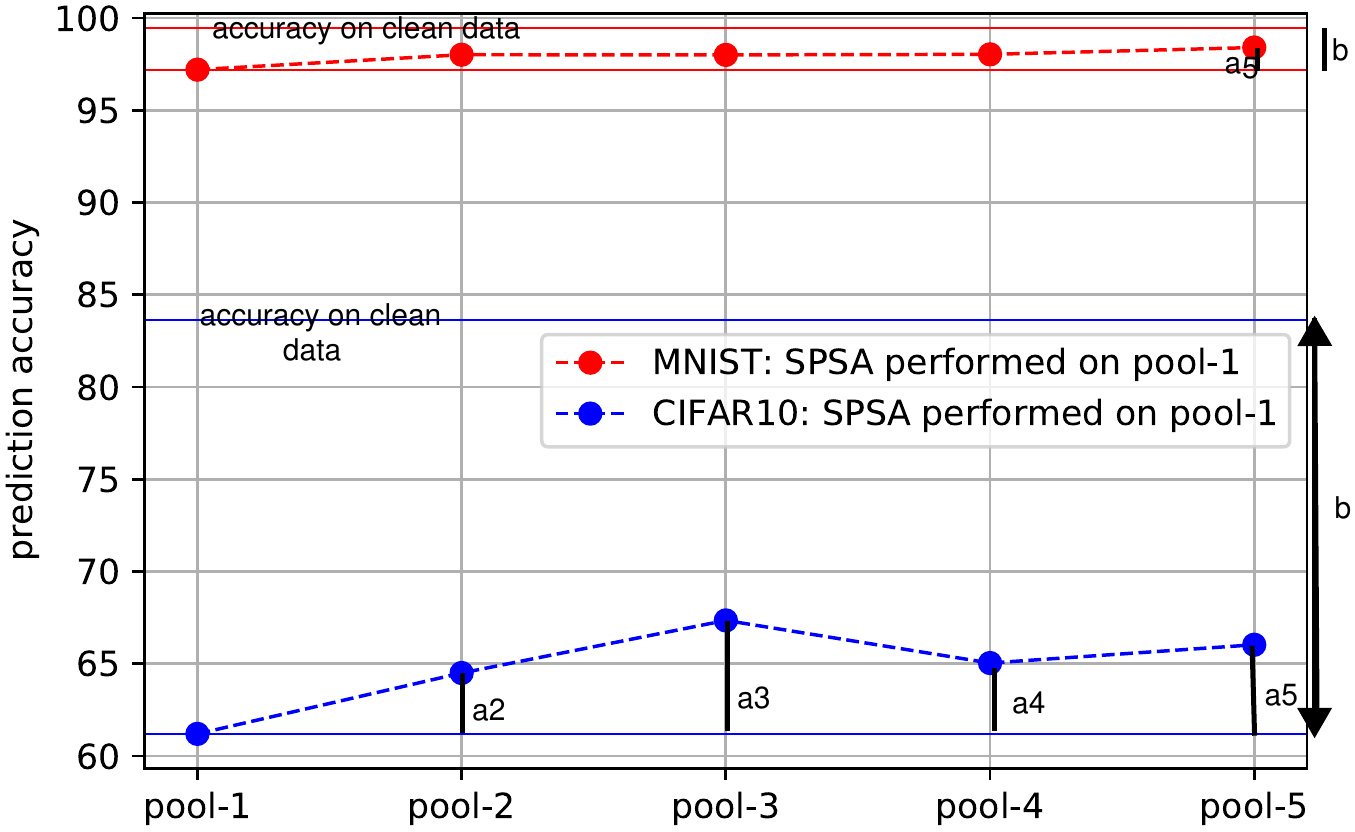}
   \vspace*{-1em}
    \caption{Impact of model pool renewal on repeating previously successful SPSA queries: Prediction accuracy of pools 2-5 of models on adversarial examples is generated through multiple queries on pool-1.}
    \label{fig:repeated}
    \vspace*{-2em}

\end{figure}

\vspace{-1em}

    
 


    
 


 
 \begin{figure*} 
    \centering
  \subfloat[$p$ vs. accuracy.\label{fig:p-vs-acc}]{%
       \includegraphics[width=0.49\linewidth]{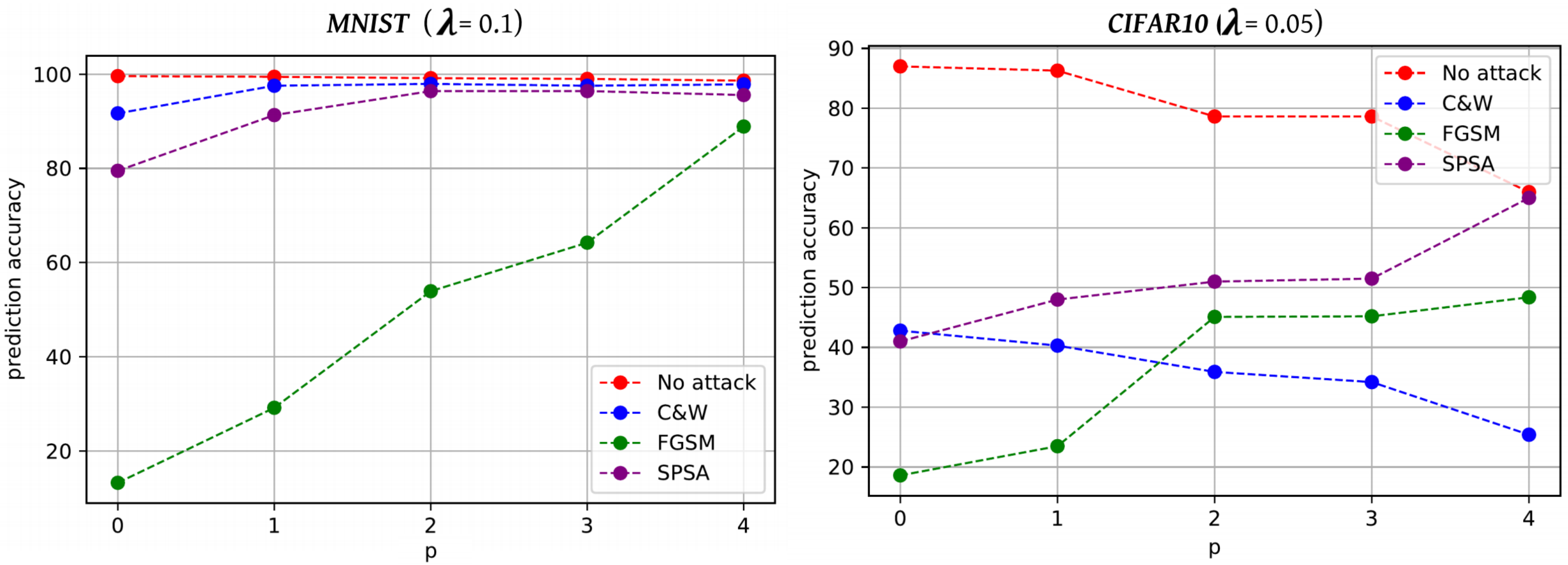}}
    \hfill
  \subfloat[$p$ vs. average transferability rate.\label{fig:p-vs-transf}]{%
        \includegraphics[width=0.49\linewidth]{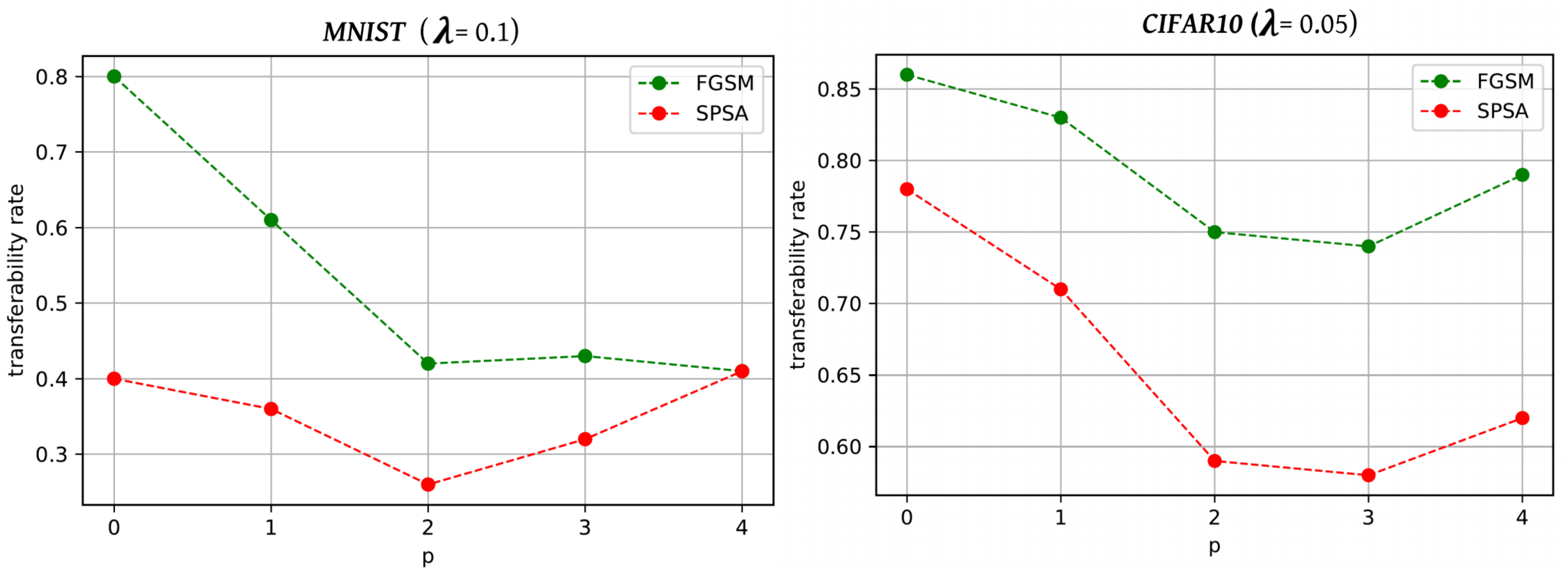}}
 \vspace{-0.5em}
  \caption{\# adversarially trained models $p$ with respect to accuracy (left) and transferability rate (right).}
   \label{fig:p-vs-acc-trans} 
   \vspace{-1.5em}
\end{figure*}

 \subsection{Model Pool Renewal vs. Repeated Attacks}
\label{mtd_impact}
To diagnose the impact of the model pool renewal on the effectiveness of repeated adversarial queries, we perform the SPSA attack by querying only pool-1 of \sysname{}. Then we test the generated adversarial examples on the ulterior pools of models (i.e., pools 2--5). For this experiment, we adopt the notation \textit{Failed Repeated Queries (FRQ)} that represents the number of ineffective repeated adversarial queries ($``a"$ in Figure \ref{fig:repeated}) from the total number of  repeated previous adversarial queries ($``b"$ in Figure \ref{fig:repeated}). Results are shown in Figure \ref{fig:repeated}. With respect to the baseline evasion results (i.e., accuracy of pool-1 on SPSA), we observe for both datasets an increase of the accuracy (hence robustness) of ulterior pools (i.e., pools 2--5) on SPSA data generated by querying pool-1. These findings indicate that some of the adversarial examples that were successful on pool-1 are not successful on ulterior pools (i.e., \textit{FRQ}$>0$). More precisely, on average across different pools, $ \approx 87\%$ of the previously effective adversarial examples failed to fool ulterior pools on MNIST ($FRQ = \frac{a}{b} \approx 87\%$). As for CIFAR10, $\approx 21\%$  of previously effective adversarial examples are not successful on ulterior pools ($FRQ = \frac{a}{b} \approx 21\%$). These results reveal the impact of the \textit{model pool renewal} on defending against repeated adversarial queries. However, we note that unlike MNIST, repeated CIFAR10 adversarial queries are more likely to continue to be effective on ulterior pools ($\approx 79\%$ are still effective), which indicates the high transferability rate of SPSA examples across different pools.

\noindent \fbox{\parbox{.96\linewidth}{
{\small
\textbf{Observation 4:} \sysname{} significantly limits the success of \textit{repeating previously successful adversarial examples} due to the \textit{model pool renewal} step that regularly and seamlessly updates the model pool to invalidate patterns in adversary's observations}.}}


\vspace{-1em}
\subsection{Impact of Model Pool Generation Components}
\label{eval:components}
We now focus on the impact of each component of the student model generation steps on \sysname{}'s \textit{robustness} and \textit{transferability rate across student models}. To that end, we generate different pools of $n=4$ student models using $0<\lambda<\lambda_{max}$ and $0\leq p \leq n$.
We monitor changes in \textit{accuracy} and {\em ATR} across the $n$ student models for different values of $\lambda$ until the maximum bound $\lambda_{max}$ is reached.
Additionally, we perform a similar experiment where we try all different possible values of $0\leq p \leq n$. Finally, we evaluate the effectiveness of training each student model on a distinct set and its impact on the reduction of the {\em ATR} compared to using the same $X_{train}$ to train all student models.

\textbf{Retraining on adversarial data.}
For this experiment, we fix a $\lambda$ value that offers acceptable model accuracy and try all possible values of $p$ (i.e., $0\leq p \leq 4$).

\textit{Impact on robustness against adversarial examples:}
Figure \ref{fig:p-vs-acc} shows accuracy of \sysname{} with respect to $0\leq p \leq 4$, for both datasets. For all attacks on MNIST, a $0\rightarrow n$ increase in $p$ leads to a higher robustness. Similar results are observed for SPSA and FGSM on CIFAR10. However, the accuracy on C\&W data is lower when $p$ is higher which is consistent with CIFAR10 results of \sysold{} in Table \ref{tab:evasion_res} ($p=8$ vs. $p=9$). As stated earlier (Section \ref{eval:robust}), adversarial training on CIFAR10 leads to a comparatively higher accuracy loss on clean test data. Consequently, the accuracy on clean data decreases when we increase $p$ (``No attack" in Figure \ref{fig:p-vs-acc}). We conclude that, retraining student models on adversarial data is a crucial step to improve the robustness of \sysname{}. However, $p$ needs to be carefully picked to reduce accuracy distortion on clean data caused by adversarial training (especially for CIFAR10), while maximizing \sysname{}'s robustness. From Figure \ref{fig:p-vs-acc}, for both datasets, $p=3$ serves as a practical threshold for $n=4$ (it balances the trade-off between reducing accuracy loss on clean data and increasing accuracy on adversarial data). 

For $p=0$, although no student model is adversarially trained, we observe an increase in \sysname{}'s robustness. For instance, compared to the robustness of the undefended model on MNIST reported in Table \ref{tab:evasion_res}, despite $p=0$ (Figure \ref{fig:p-vs-acc}), the accuracy using a pool of $4$ models is improved on FGSM data ($9.98\% \rightarrow \approx 18\%$). Similar results are observed for C\&W on both datasets (MNIST: $0\% \rightarrow \approx 91\%$, CIFAR10: $0\% \rightarrow \approx 42\%$). These findings, once again, indicate the impact of the MTD aspect on increasing \sysname{}'s robustness against evasion attacks.

\textit{Impact on the evasion transferability across student models:}
Adversarially training a subset of student models leads to more diverse student models compared to those trained on just clean data -- this might reduce {\em ATR} across models.  Figure \ref{fig:p-vs-transf} shows {\em ATR} of SPSA and FGSM adversarial examples across student models for $0\leq p \leq 4$. We choose SPSA and FGSM in view of their high transferability across ML models. Figure \ref{fig:p-vs-transf} shows that, for both datasets, {\em ATR} of both attacks is at its highest rate when $p=0$. Then, it decreases for larger values of $p$ (i.e., $p=1\rightarrow 3$), until it reaches a minimum (i.e., $p=2$ for MNIST and $p=3$ for CIFAR10). Finally, {\em ATR} of both attacks increases again for $p=n=4$. In this final case, all student models are adversarially trained, therefore they are less diverse compared to $p \in \{1,2,3\}$. We conclude that the choice of $p$ has an impact, not only on the overall performance of \sysname{}, but also on the {\em ATR} across student models. 

\begin{figure*} 
    \centering
  \subfloat[$\lambda$ vs. accuracy.\label{fig:lambda-vs-acc}]{%
       \includegraphics[width=0.49\linewidth]{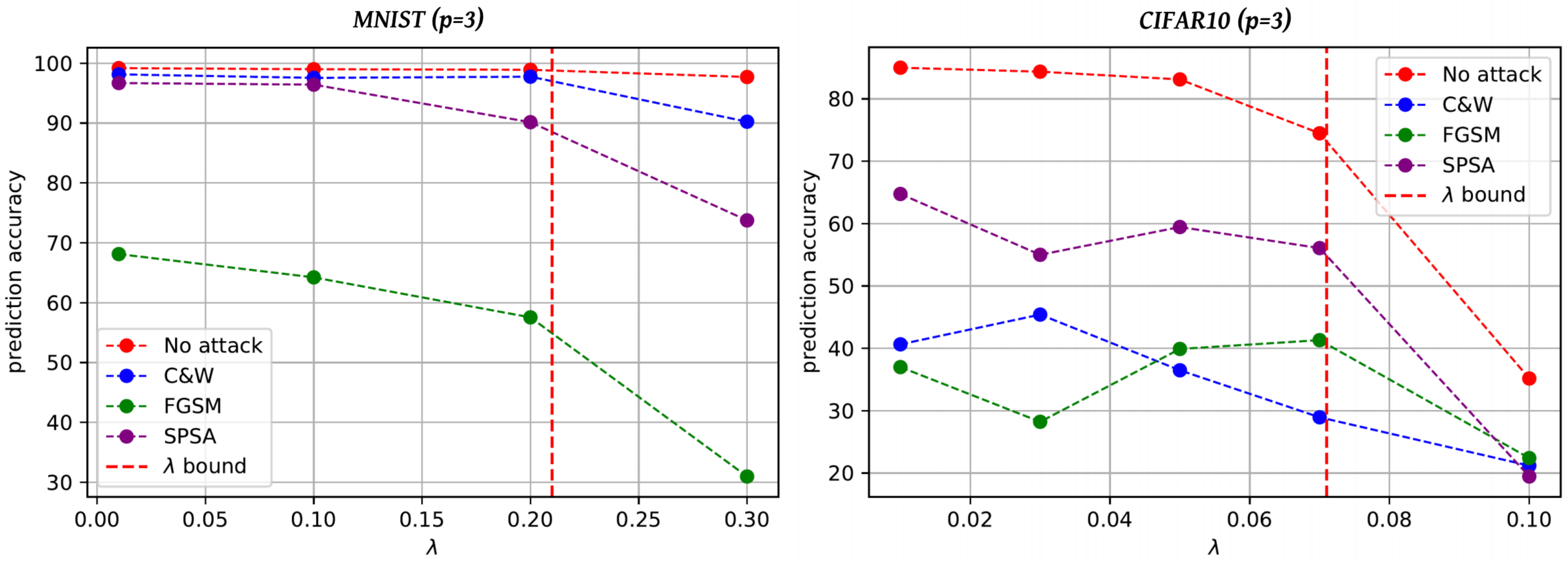}}
    \hfill
  \subfloat[$\lambda$ vs. average transferability rate.\label{fig:lambda-vs-transf}]{%
        \includegraphics[width=0.49\linewidth]{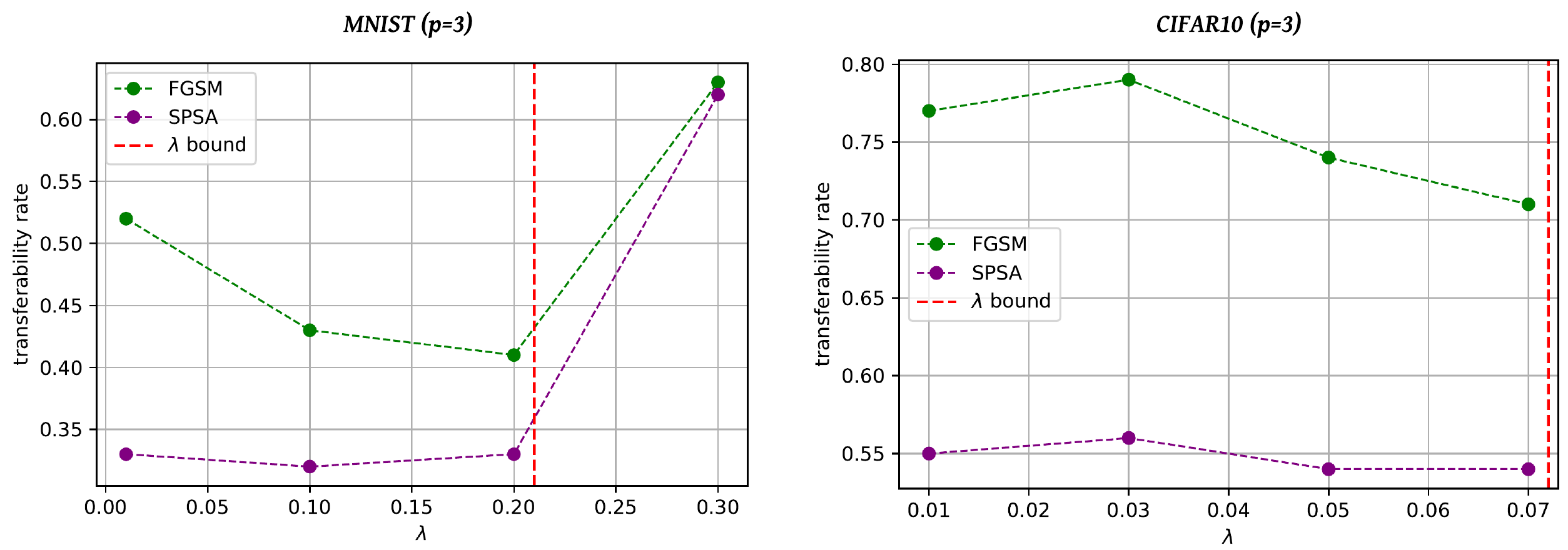}}
 \vspace{-0.5em}
  \caption{Noise scale $\lambda$ vs. accuracy (left) and average transferability rate (right).}
   \label{fig:lambda-vs-acc-trans} 
   
   \vspace{-1.5em}
\end{figure*}

\textbf{Noise Scale $\lambda$.}
We begin with $p=3$ and incrementally try different configurations of $\lambda>0$ until we reach a maximum bound $\lambda_{max}$. In Figures \ref{fig:lambda-vs-acc} and \ref{fig:lambda-vs-transf}, with respect to different values of $0< \lambda< \lambda_{max}$, we investigate the impact of the {\em weights perturbation} step on accuracy and on {\em ATR} across student models. For both datasets, $\lambda_{max}$ is presented as a vertical bound (i.e., red vertical line).

\textit{Impact on ATR:} Figure \ref{fig:lambda-vs-transf} indicates that an increase in the noise scale $\lambda$, generally, leads to the decrease of the transferability rate of adversarial examples across student models, for both datasets. This observation is consistent with our intuition (in Section \ref{sec: approach}) that higher distortions on $f_b$ weights lead to the generation of more diverse student models.

\textit{Impact on the accuracy:} For MNIST dataset, we observe that higher model weights distortion (i.e., higher $\lambda$) leads to less performance on clean data and on all the studied adversarial data (e.g., C\&W, FGSM and SPSA). As for CIFAR10, Figure \ref{fig:lambda-vs-acc} indicates different results on SPSA and FGSM. For instance, unlike C\&W examples, which are much less transferable, the accuracy on FGSM data reaches its highest when $\lambda=0.05$. Similarly, we observe an increase of the accuracy on SPSA data for $\lambda=0.05$. Next, we further discuss the difference in accuracy patterns on FGSM and SPSA between MNIST and CIFAR10.




\textit{Best $\lambda$ configuration:} On MNIST, $\lambda=0.1$ represents a tradeoff that balances the reduction of the tansferability rate and the reduction of the accuracy loss, which are conflicting. As for CIFAR10, we choose $\lambda=0.05$ to balance between \sysname{} performance against non-transferable attacks (e.g., C\&W) and transferable attacks (e.g., FGSM and SPSA).

    
 


    
 



\noindent \fbox{\parbox{.96\linewidth}{
{\small
\textbf{Observation 5:} \sysname{}'s performance is influenced by the values of its hyper-parameters (e.g., $\lambda$, $n$ and $p$). Empirically estimating the optimal configuration of \sysname{} contributes to the reduction of the {\em ATR} across student models, which leads to an increased robustness against adversarial examples}.}}

\begin{table}[t!]

\centering
  \scalebox{.88}{
   \begin{tabular}{|l!{\color{black}\vrule}l!{\color{black}\vrule}l!{\color{black}\vrule}!{\color{black}\vrule}!{\color{black}\vrule}l!{\color{black}\vrule}l!{\color{black}\vrule}l!{\color{black}\vrule}l!{\color{black}\vrule}}
       \hline
       
&  \multicolumn{2}{c!{\color{black}\vrule}}{\textbf{MNIST-CNN}}
&  \multicolumn{2}{c!{\color{black}\vrule}}{\textbf{CIFAR10-CNN}}\\
       \hline
       & \textbf{FGSM} &  \textbf{SPSA} & \textbf{FGSM} &  \textbf{SPSA}\\ 
       \hline
       Retraining on $X_{train}$ & 0.88 & 0.52&0.95&0.84\\ 
       \hline\hline
       Retraining on $T_i(X_{train})$  &\textbf{0.80} &\textbf{0.40}&\textbf{0.86}&\textbf{0.78}\\ 
       \hline
      
   \end{tabular}}
 
\caption{Comparison of {\em ATR} of FGSM and SPSA when student models are retrained on $X_{train}$ vs. on $T_i(X_{train})$ ($p=0$).}
\label{tab:transf}
\vspace{-4em}
\end{table} 
  
\textbf{Retraining on transformed data:}
We now evaluate to what extent using data transformation reduces the transferability rate of adversarial examples. To that end, we compute the {\em ATR} of FGSM and SPSA, and we compare it with the baseline case where all student models are retrained on the same training set $X_{train}$. To precisely diagnose the impact of data transformations on transferability, we exclude the effect of Step-3 by using $p=0$, in addition to the same $\lambda$ configuration adopted before. Therefore, we note that the following results do not represent the actual transferability rates of \sysname{} student models (covered in previous discussions). Results reported in Table \ref{tab:transf} show that performing different data transformations $T_i$ on the training set $X_{train}$ before retraining leads to more diverse student models. For instance, we observe $\approx -8,5\%$ less transferable FGSM examples on average across both datasets and an average of $\approx -9\%$ less transferable SPSA examples. 

\noindent \fbox{\parbox{.96\linewidth}{
{\small
\textbf{Observation 6: } \textit{Training data transformation} is effective at reducing {\em average transferability rate}}.}}

Despite \sysname{} advances to reduce ATR, the transferability challenge still has room for improvement. Prior work examined the adversarial transferability phenomenon \cite{transferability16, WhyTransfer19}. Yet, rigorous theoretical analysis or explanation for transferability phenomenon is a work in-progress in the literature.
\vspace{-2em}
\section{Conclusion}\label{sec: concl}
\sysname{} advances \sysold{} by overhauling the model scheduling strategy via OOD detection. In \sysname{}, model weights perturbation, data transformation, adversarial training, OOD-powered dynamic model pool scheduling, and seamless model pool renewal work in tandem to thwart  adversarial examples. 
Our comprehensive experiments across white-box and black-box attacks on benchmark datasets suggest \sysname{} significantly outperforms adversarial training and improves \sysold{} on robustness to adversarial examples. As in \sysold{}, it does so while preserving accuracy on clean data and reducing attack transferability among models in the \sysname{} pool.
\vspace{-1em}

\bibliographystyle{IEEEtran}
\bibliography{main}

\end{document}